\documentclass[12pt]{amsart}

\usepackage{amsfonts,amssymb,amsthm}
\usepackage{epsfig,graphicx}
\usepackage{mathptmx}      % use Times fonts if available on your TeX system
\makeatletter

\renewcommand \thesection {\@arabic\c@section}
\renewcommand\thesubsection   {\thesection .\@arabic\c@subsection}
\renewcommand\thesubsubsection{\thesubsection .\@arabic\c@subsubsection}
\renewcommand\theparagraph    {\thesubsubsection .\@arabic\c@paragraph}
\makeatother

\newcounter{subequation}
        {\addtocounter{equation}{-1}%
        \stepcounter{subequation}%
        \begin{equation}}%
        {\end{equation}%
}

\newcommand\abs[1]{{| #1 |}}

\newcommand\norm[1]{{\| #1 \|}}

\newcommand{\rmb}{{\, \mathrm b}}
\newcommand{\rmd}{{\, \mathrm d}}
\newcommand{\rme}{{\, \mathrm e}}
\newcommand{\rmi}{{\mathrm i}}

\newcommand{\bfU}{{\mathbf U}}

\newcommand{\cA}{{\mathcal A}}

\newcommand{\cF}{{\mathcal F}}

\newcommand{\cN}{{\mathcal N}}

\newcommand{\CC}{{\mathbb C}}
\newcommand{\EE}{{\, \mathbb E \, }}

\newcommand{\NN}{{\mathbb N}}
\newcommand{\PP}{{\mathbb P}}

\newcommand{\RR}{{\mathbb R}}

\newcommand{\TT}{{\mathbb T}}
\newcommand{\ZZ}{{\mathbb Z}}

\newtheorem{defn}{Definition}[section] % This defines the counter for
\newtheorem{Assms}[defn]{\sc Assumptions}

\newtheorem{Prop}[defn]{\sc Proposition}
\newtheorem{Rema}[defn]{\sc Remark}
\newtheorem{Theo}[defn]{\sc Theorem}

\begin{document}

\title{Gaussian convergence for stochastic acceleration of $\cN$ particles 
       in the dense spectrum limit}

% \titlerunning{Stochastic wave field acceleration}       
                   % if too long for running head : J. Stat. Phys. 

\author{Yves Elskens}
      \address{Equipe turbulence plasma, case 321, 
      PIIM, UMR 7345 CNRS,
       Aix-Marseille universit\'e,
       campus Saint-J\'er\^ome, FR-13397 Marseille cedex 13 \\
       {\tt{yves.elskens@univ-amu.fr}}
 }
%\date{Received: date / Accepted: date}
\date{July 9, 2012}
\maketitle

\begin{abstract}\noindent
     The velocity of a passive particle in a one-dimensional wave field 
     is shown to converge in law to a Wiener process, 
     in the limit of a dense wave spectrum
     with independent complex amplitudes, where the random phases distribution 
     is invariant modulo $\pi/2$ and the power spectrum expectation is uniform. 
     The proof provides a full probabilistic foundation
     to the quasilinear approximation in this limit. 
     The result extends to an arbitrary number of particles, 
     founding the use of the ensemble picture for their behaviour 
     in a single realization of the stochastic wave field.
  \\
%  ***
  \\ %
%    \PACS{      % for J. Stat. Phys. 
   \\ %   
%  \subclass{% for J. Stat. Phys. 
 MSC :  { 
      34F05, % \and 
      60H10, % \and 
      82C05, %  \and 
      82D10, % \and 
      60J70, % \and 
      60K40 }
%  \\
%     MSC~2000~:
%     34F05  Equations and systems with randomness \\
%     60H10  Stochastic ordinary differential equations \\
%     60J70  Applications of diffusion theory \\
%     82C05  Classical dynamics and nonequilibrium statistical mechanics (general) \\
%     82D10  Plasmas \\
%     60K40  Other physical applications of random processes
%     \\ *** \\
%     PACS~:
%        05.45.-a  Nonlinear dynamics and nonlinear dynamical systems
%  \\    41.75.-i  Charged particle beams
%  \\    52.35.-g  Waves, oscillations, and instabilities in plasmas and intense beams
%  \\    84.40.-x  Radiowave and microwave (including millimeter wave) technology
%  \\    29.27.-a  Beams in particle accelerators
%  \\
%  ***
\end{abstract}

%\begin{keyword}[class=AMS]
%\kwd[Primary ]{}
%\kwd{}
%\kwd[; secondary ]{}
%\end{keyword}

%\begin{keyword}
%\kwd{}
%\kwd{}
%\end{keyword}

\vfill \hrule
\smallskip
\noindent \textbf{preprint submitted for publication} \\
%  DRAFT -- private communication, not for distribution
  % To appear in: \textit{Journal of Statistical Physics}.
  \par
\noindent \copyright{2012} The author. Reproduction of this article, in its entirety,
for noncommercial purposes is permitted.

\newpage

%===========================================================
\section{Introduction}
 \label{secIntro}
%===========================================================

We first recall the physical setting in sec.~\ref{secPhySet}. 
Indeed the quasilinear approximation is an ubiquitous scheme for deriving 
irreversible, diffusion-like equations from many-body dynamics, 
involving a ``propagation of chaos'' kind of argument in a system with mean-field behaviour. 
Original derivations of this approximation are sketched in sec.~\ref{secOriQL}.
A different, more recent analysis in the framework of hamiltonian chaotic dynamics is recalled in
sec.~\ref{secHamilton}.
As these arguments are well detailed in the literature, 
we do not reproduce the calculations and proofs 
but merely highlight their key points. 
Our main result and particulars of the present work 
are stressed in sec.~\ref{secThisWork}.

%  *** explain ``dense spectrum = ???''

%===========================================================
\subsection{Physical setting}
 \label{secPhySet}
%===========================================================

The motion of a particle in the field of many waves is a fundamental process in
collisionless plasma physics. Even if the particle motion does not feed back on 
the wave parameters, viz.\ for a test particle, undergoing passive transport,
this problem still presents open issues. 
Its most elementary instance, in one space dimension, is
also a benchmark for approximation techniques.

This one-dimensional problem describes the motion of a particle in a longitudinal, 
electrostatic, time-dependent potential. 
Electrostatic modes occur in various contexts \cite{GoRu95,DoMaAuh05}, including
(i) the non-relativistic regime of Coulomb plasmas, where magnetic fields are negligible~; 
(ii) particle motion parallel to the applied magnetic field in strongly magnetized plasmas~;  
(iii) particle motion along the axis of a waveguide, such as traveling wave tubes 
used as amplifiers in telecommunications.
The time dependence of the field leads to the propagation of waves, which are longitudinal~:
Langmuir waves are the simplest collective modes in plasmas.
When it applies (in particular for hot plasmas), 
the neglect of collisions in particle dynamics within plasmas 
rests on the long-range nature of Coulomb interaction 
leading to a mean-field picture in both the Vlasov kinetic equation 
and the Euler fluid models.  

In many situations, the wave field evolution involves a response to the particle motion.
However, in some instances the particle feedback on the electrostatic field is negligible, and one may take the field as given.
The equations of motion for a particle with charge $\mathfrak e$ and mass $\mathfrak m$
then read
\begin{eqnarray}
  \frac{\rmd X}{\rmd t}
  & = &
  V(t)
  \label{Nq1}
  \\
  \frac{\rmd V}{\rmd t}
  & = &
  \frac{\mathfrak e}{\mathfrak m}\  E(X(t), t)
  \label{Np1}
\end{eqnarray}
where the electric field $E$ is a prescribed process. 
It is convenient to represent $E$
as a Fourier series, $E(x,t) = \sum_m E_m \rme^{\rmi (k_m x - \omega_m t + \varphi_m)}$,
where amplitudes $E_m$ and phases $\varphi_m$ may be tunable, while $(k_m, \omega_m)$
are given by the waves
  dispersion relation.\footnote{Note that this model differs
      from stochastic acceleration problems in a random potential,
      for which the field $E(x,t)$ reduces to a static random $E(x)$.}
The key effect of a single wave with  phase velocity $v_{\varphi, m} = \omega_m/k_m$ 
on a particle with velocity $v$ is a tendency \cite{DovEsMa05} 
to reduce the relative velocity $\abs{v - v_{\varphi, m}}$, 
and this effect works best when 
${\mathfrak m} (v - v_{\varphi, m})^2 \sim \abs{{\mathfrak e} E_m / k_m}$.
The competition between two waves $m, m'$ in attempting this synchronization 
is measured by the wave overlap parameter 
\begin{equation}
  s_{m,m'} 
  := 
  2 \abs{\frac{\mathfrak e}{\mathfrak m} } ^{1/2} \ 
      \frac{ \abs{E_m / k_m}^{1/2} + \abs{E_{m'} / k_{m'}}^{1/2} }
              {\abs{v_{\varphi, m} - v_{\varphi, m'}}} \, . 
\label{defs}
\end{equation}
When this parameter is small, 
a particle cannot interact strongly with both waves simultaneously, 
and the dynamics can be analyzed perturbatively. 
Actually, the dynamics (\ref{Nq1})-(\ref{Np1}) is well known  to be nonintegrable 
as soon as there is more than a single wave phase velocity~; 
the two-wave model is a paradigm of hamiltonian chaos with 1.5 degrees of freedom, 
with a KAM limit as $s \to 0$, and transition to ``large scale chaos'' as $s \gtrsim 1$. \cite{Escande85,EEbook}

Denoting by
$\Delta v_\varphi$ the typical relative velocity of a wave with respect to its
nearest neighbours, by $k_{\mathrm{typ}}$ a typical wavevector, 
and by $E_{\mathrm{typ}}$ a typical amplitude, 
the regime of interest for this paper is the \emph{dense spectrum limit}, 
where a particle is typically influenced significantly by many waves~;
in this regime the typical resonance overlap parameter 
$s 
    = 4 \sqrt{\abs{({\mathfrak e}  E_{\mathrm{typ}}) / ({\mathfrak m} k_{\mathrm{typ}}) }}%^{1/2} 
    / \Delta v_\varphi$
is large. 
It is then tempting to consider the acceleration in the
right hand side of (\ref{Np1}) as an approximate white noise, 
and the particle velocity $V$ as a kind of
diffusion process~: this is the core of the quasilinear approximation
\cite{Romanov61,Vedenov61,Vedenov62,Drummond62}. The latter is widely used, 
in diverse physical contexts, as it is easily
implemented and relies on simple ideas, which we comment in the following. 
% However, there is a gap between magnitude
% estimates and actual convergence proofs.

%===========================================================
\subsection{Original derivations of the quasilinear approximation}
 \label{secOriQL}
%===========================================================

Classical derivations of the quasilinear approximation in plasma physics textbooks,
e.g.\ \cite{Kadomtsev65,GoRu95,HaWa04}, start from viewing the motion of the test particle 
as the transport of a measure $\rmd \mu = f \rmd x \rmd v$ on $(x,v)$ space 
(with $f(.,.,t)$ possibly a distribution),
\begin{equation}
  \partial_t f + v \, \partial_x f
  =
  - \frac{\mathfrak e}{\mathfrak m}\  E(x, t) \, \partial_v f
\label{Vlasov} \, ,
\end{equation}
and begin an iterative solution with respect to $E$, 
\begin{eqnarray}
  f(x,v,t)
  &=&
  f(x-vt, v, 0) 
  - \frac{\mathfrak e}{\mathfrak m}  \int_0^t E(x-v (t-t_2), t_2) \partial_v f((x-v (t-t_2), v, t_2) \rmd t_2
  \cr
  && % \quad
  + (\frac{\mathfrak e}{\mathfrak m})^2  
      \int_0^t \int_0^{t_2} E(x-v (t-t_2), t_2) E(x-v (t_2-t_1), t_1)  
      \cr && \hskip4cm 
      \partial_v f((x-v (t-t_1), v, t_1)    \rmd t_1 \rmd t_2
\label{VlasovIter}  \, .
\end{eqnarray}
On performing an $x$ average which highlights
the correlation function of the electric field $E$, 
one then relies on independence of $E$ from the
(slaved, passive, tracer) particle distribution $\rmd \mu$ to eliminate the first order term, and
obtains an integro-differential evolution equation for
the $x$-averaged $\bar f$. 
Then, on considering that the velocity process $V$ must be Markov
on time scales longer than the correlation time of $E$, 
the equation for $\bar f (v,t)$ reduces to
\begin{equation}
  \partial_t \bar f
  -
  \partial_v ( D(v) \, \partial_v \bar f )
  =
  0
  \label{diffeq}
\end{equation}
where the velocity-dependent diffusion coefficient
\begin{equation}
  D(v)
  =
  \frac{{\mathfrak e}^2}{{\mathfrak m}^2} \int_0^\infty
    \langle E(x, t) E(x-v\tau, t-\tau) \rangle \, \rmd \tau
  =
  \frac{\pi {\mathfrak e}^2}{{\mathfrak m}^2}
    \int \delta(\omega - kv) \langle \abs{E_k}^2 \rangle \, \rmd k
  \label{Dcoef}
\end{equation}
is determined by the wave field lagrangian autocorrelation, with appropriate averaging
$\langle \cdot \rangle$ and assuming that phases $\varphi_m$ are independent and
uniformly distributed. % (the ``random phase approximation''). 
The Fourier form in
(\ref{Dcoef}), with a Dirac distribution, obtains in the continuous spectrum limit. 

The ``appropriate averaging'' $\langle \cdot \rangle$ may imply 
(see e.g.\ sec.~9.4 in \cite{HaWa04}) that one no longer considers 
the evolution of test particles velocity distribution $\int f(x,v,t) \rmd x$ 
in a single realization of the field $E$ but rather the \emph{expectation} 
of this $\int f(x,v,t) \rmd x$ with respect to the ensemble of wave fields. 
Such a view pertains to the statistics of particle velocities 
collected from repeated experiments, but it does not apply a priori 
to the description of transport in a single realization, 
as stressed in more general terms e.g.\ p.~45 in \cite{HaWa04}. 

This derivation may be criticized (within its own viewpoint) 
on the ground that, however small the coefficient $E$
may be, the differential operator $\partial_v$ is unbounded for many function spaces.
Formal, diagrammatic \cite{Bourret62a,Bourret62b,Bourret65,Thomson73,Brissaud74} expansions 
in $E \, \partial_v$ are therefore less straightforward than they may seem.

An alternative derivation, based on particle motion and $E$-power expansion, also leads to
the diffusion equation (\ref{diffeq}) via its Langevin counterpart, 
assuming that the particle velocity is a Markov
process and computing the first two moments of its increments \cite{Sturrock66}.
In this context, the ``random phase approximation'' is actually invoked so that,
\emph{to practical ends}, 
``lagrangian'' (as seen by a test particle) phases 
$k_m (x - v (t - t_j)) +  \varphi_m - \omega_m t_j$,
can be considered as independent (uniformly distributed modulo $2\pi$) 
random variables 
for any relevant sequence of times $t_j$ and wave indices $m$, 
viz.\  not only at a single time 
(this is imposed by the very distribution of parameters $\varphi_m$)
but as if their values were ``refreshed'' repeatedly. The boldness of such an assumption,
akin to the propagation of molecular chaos in gas theory \cite{Kac56,Kac59},
fueled the debate on the validity of the quasilinear approximation 
(as a preamble to the further debate focusing on the self-consistent problem, 
where wave amplitudes and phases evolve under particle feedback)
\cite{CEV,Ishihara93a,LavalPesmePPCF}.

Mathematically, the ``repeated random phase approximation'' is valid,
under a few more technical conditions \cite{Papanicolaou74},
in the limit $\varepsilon \to 0$ after a time rescaling, $\tau = \varepsilon^2 t$, 
when the field $E$ is \emph{mixing}, in the sense that the process $E(.,t)$ is adapted 
to a family of $\sigma$-algebras $\cF_s^t$, $0 \leq s \leq t \leq \infty$, 
with $\cF_{s_1}^{t_1} \subseteq \cF_{s_2}^{t_2}$ 
for $0 \leq s_2 \leq s_1 \leq t_1 \leq t_2 \leq \infty$, 
with a probability measure $\PP$ such that the rate function
\begin{equation}
 \rho(t) 
 := 
 \sup_{s \geq 0} \sup_{A \in \cF_{s+t}^\infty , B \in \cF_0^s} 
    \abs{\PP(A|B) - \PP(A)}
\label{ratefct}
\end{equation}
satisfies the condition $\int_0^\infty \sqrt{\rho(t)} \rmd t < \infty$. 
Typical examples of such mixing processes $E$ are ergodic Markov processes on a compact space
 \cite{Papanicolaou74},
but time-periodic fields as discussed e.g.\  in Refs~\cite{CEV,BE98,EEbook,ElPa10} 
fail to meet the mixing condition.

The use of an adjoint formulation instead of trajectories is generally motivated by the
traditional viewpoint of kinetic theory, interested in following many particles (in which case,
including the self-consistent dynamics where the evolution of $E$ depends on $f$, 
measures provide a natural description, see e.g.\ ch.~I.5 in \cite{Spohn91}),
by the fact that the Vlasov and diffusion equations are linear for $f$, 
and by the familiar description of Markov processes in
terms of their generator. 
Yet a single physical realization of the wave field $E$ acts on
a particle distribution quite differently from the way an ensemble of independent
realizations would act on a single particle \cite{BE98Std}. 
The decorrelation assumption is crucial in
claiming that the ensemble may describe a single experiment with many particles.
Besides, if the Markov assumption fails, the single-time distribution function
$f(x,v,t | x_0, v_0, t_0)$ may fail to describe properly 
the joint $n$-time distribution $F(x_1, v_1, t_1 \ldots x_n, v_n, t_n)$. 
Therefore we revisit the derivation of quasilinear equations from a particle viewpoint, and
possibly reach a Markov description in an appropriate limit.

%===========================================================
\subsection{Hamiltonian dynamics approach}
 \label{secHamilton}
%===========================================================

This dynamics-based program was significantly advanced by B\'enisti and Escande
\cite{BE97,BE98}, who proved the validity of the velocity diffusion picture for the
dynamics defined by hamiltonian
\begin{equation}
  H
  =
  \frac{p^2}{2 {\mathfrak m}} + \cA \sum_{m = -M}^M \cos ( q - mt + \varphi_m)
 \label{Hbe}
\end{equation}
in the limit $M^{3/2} \gg \cA/{\mathfrak m} \to \infty$, when phases $\varphi_m$ are
independent and uniformly distributed in $[0,2\pi]$. Their derivation relies on the
strong chaos (as $s \to \infty$) in particle dynamics associated with the limit, 
and on the fact that, at a time $t$, only
waves with a phase velocity such that $\abs{v_\varphi - p(t)/{\mathfrak m}} \lesssim \Delta
v_{\mathrm b}$ act strongly (nonperturbatively) on the particle. 
Waves beyond the ``resonance box half-width'' 
$\Delta v_{\mathrm b} \sim 5 (\cA/{\mathfrak m})^{2/3}$ can
be eliminated from the dynamics (their overall statistical effect is exponentially small 
in $\abs{v_\varphi - p/{\mathfrak m}} / \Delta v_{\mathrm b}$) by a canonical transformation, 
so that the velocity process is Markov on scales wider than the resonance box. 
On the other hand, for shorter time scales, the particle velocity needs a time
of the order of unity to sample correlations associated with the discreteness of the
frequency spectrum, so that it is chaotic and wanders so much that it eventually moves to
another resonance box. 
Moreover, for short time scales, they show how to relax the assumption that all phases are independent to the requirement that any two phases influence negligibly the particle motion \cite{BE97,BE98,EEbook}.

This argument was complemented by the observation that the short-time quasilinear
approximation holds for times $0 < t \lesssim D^{-1/3} \ln s$, and that the Markov
approximation holds for times $t \gtrsim D^{-1/3}$  \cite{EE02,EEbook} 
($D^{-1/3}$ is also related to the Lyapunov time scale for the 
divergence of microscopic trajectories in a typical wave field  \cite{EEbook}), 
so that the quasilinear approximation holds for all times in the dense spectrum limit $s \to \infty$.

For technical simplicity, the hamiltonian model (\ref{Hbe}) involves three restrictions 
with respect to the original dynamics (\ref{Nq1})-(\ref{Np1}) : 
all amplitudes are equal, all wavevectors are equal, 
and all phase velocities are equally spaced. 
B\'enisti and Escande \cite{BE98} sketch 
how their arguments can be extended to lift these restrictions. 
The hamiltonian (\ref{Hbe}) also stresses the spectrum discreteness time scale, 
as $\Delta v_\varphi = 1$, 
which can generate correlations over long times \cite{BE97,BE98Std}.

%===========================================================
\subsection{Position of this work}
 \label{secThisWork}
%===========================================================

In the present work we extend the approach initiated in \cite{Els07,ElPa10}
and revisit  the B\'enisti--Escande result with the language of probability theory. 
We express the wave field as a sum of $N \to \infty$ independent components per unit frequency interval, so that the overlap parameter $s$ diverges in the limit $N \to \infty$.
We first prove in Theorem~\ref{ThmU} that, in the resulting dense wave spectrum limit $s \to \infty$, 
the wave field acting on a particle for $0 \leq t \leq 2\pi$ converges in law to the field 
associated with a ``white noise''. 
This enables us to derive Proposition~\ref{WienerOnePart} 
and Theorem~\ref{WienerNPart} on particle motion.

Proposition~\ref{WienerOnePart} shows that, for $M \to \infty$,
for fixed wave power spectral density with $N \to \infty$ so that $s \to \infty$, 
the velocity of a single particle in the wave field converges in law to a Wiener process
over the time interval $[0,2\pi]$. 
While B\'enisti and Escande emphasize a hamiltonian dynamical system
approach, we focus on the velocity and express our limit theorem as a convergence
in distribution result, following essentially from central limit averages on the wave field. 
The convergence in distribution was clearly understood in \cite{BE97,BE98},
in particular through the statement that the influence of waves outside a resonance box 
is only perturbative on the statistical properties of the dynamics (p.~914 in \cite{BE98}).
The focus on $v$ is also central to the arguments in \cite{EEbook} 
which involve the characteristic function $\Phi(\gamma,t) := \EE \exp(\rmi \gamma (v(t) - v(0)))$.

We also pay attention to the behaviour of an arbitrary number $\cN$ of particles
moving in the same wave field. Their evolutions 
are not independent processes for finite $\cA$, so that
the diffusion equation (\ref{diffeq}) does not describe the evolution of the empirical
distribution $\cN^{-1} \sum_{\ell = 1}^\cN \delta (v - p_\ell/{\mathfrak m})$ ; 
this was stressed in \cite{BE98Std}. 
However, our main result, Theorem \ref{WienerNPart}, proves that, 
in the limit $s \to \infty$, particles do diffuse independently, 
even in the same wave field\footnote{The contrast 
   between this conclusion and B\'enisti--Escande's \cite{BE98Std}
   might be attributed to the asymptotic nature of our result, as $s \to \infty$. 
   We do not provide estimates for the ``convergence rate'' 
   of the empirical distribution to its Fokker-Planck limit.}
-- in agreement with the view that they generally are in different resonance boxes. 
Thereby we extend to a broad class of wave fields the conclusion of  \cite{ElPa10},
which assumed a wave field generated by Wiener processes
(viz.\ the fields obtained in the dense spectrum limit). 
This result provides some support to the traditional view that a single
realization may, in some cases, be approximated by an ensemble. 

This paper is organized as follows. We state our assumptions on
the wave field in section~\ref{secAssume}. 
%.
These enable a fast proof of the convergence of elementary processes
associated with the wave field in section~\ref{secWaveConv}.
%.
Thanks to Ref.~\cite{ElPa10} and the continuous mapping theorem \cite{Kal01}, 
the convergence of the
$\cN$-particle velocity process to the diffusion limit follows immediately in
section~\ref{secParticle}.
%.
We stress the interpretation of our techniques and results 
in section~\ref{secInterpretation}.
%.
We conclude with a discussion of open issues.

%===========================================================
\section{Wave field assumptions}
 \label{secAssume}
%===========================================================

A random variable (r.v.) $\alpha$ is symmetric \cite{Kah85,Kal01}
if $\alpha$ and $- \alpha$ have
identical distributions. Then $\EE \alpha^k = 0$ for odd $k$ if the expectation exists. 
\begin{Assms}[S2, S4]
\label{condS4}
  Given $M \in \NN = \{0, 1, 2, \ldots\}$ and $N \in \NN_0 = \{1, 2, \ldots\}$, 
  consider $(2M+1)N$ complex random variables
  $\alpha_{m,n} = A_{m,n} \rme^{\rmi \varphi_{m,n}}$.
  We say that the $\alpha_{m,n}$'s meet assumptions (S2) if
  \begin{enumerate}
    \item{they are independent and symmetric,}
    \item{$\EE A_{m,n}^2 = 1$,}
    \item{$\sup_{m,n} \EE A_{m,n}^4 \leq C_4$ for some $C_4>1$.}
  \end{enumerate}
  We say that they meet assumptions (S4) if, moreover,
  the r.v.\ $\alpha_{m,n}^2$ is also symmetric.
\end{Assms}
The additional condition
for (S4) may be called ``\emph{four-symmetry}'' for the r.v.\ $\alpha$. Examples are (i) the
r.v.\ $\rme^{\rmi (c + K \pi/2)}$, with fixed $c$ and $\PP(K=k) = 1/4$ for $k \in \{1,
2, 3, 4\}$, (ii) an isotropic complex r.v., viz.\ $\alpha = A \rme^{\rmi \varphi}$ such
that $\varphi$ is uniform on $[0, 2\pi]$ (which corresponds to a Steinhaus sequence
$\varphi_{m,n}/(2 \pi)$ \cite{Kah85}) and independent from $A$, 
and in particular (iii) a standard normal complex r.v.\
(isotropic, with exponentially distributed $A^2$).

\begin{Rema}
  The third condition in (S2) is unnecessarily stringent
  (though being met for many physical cases),
  and could be relaxed to a Lindeberg-type condition.
\end{Rema}

Occasionally we identify $\RR^2$ with $\CC$ to minimize the amount of notations. We
denote by $B$ the standard brownian motion in $C(\RR^+, \RR)$ and by $W$ the standard
brownian motion in $C(\RR^+, \CC)$, so that $B$, $\sqrt{2}\, \Re W$ and $\sqrt{2}\, \Im
W$ are independent and identically distributed (i.i.d.).

%===========================================================
\section{Convergence of the controlling wave processes}
 \label{secWaveConv}
%===========================================================

Given  $N$ real parameters
  $\sigma_n \in [0, 1]$ ($1 \leq n \leq N$),
we first introduce the $N$ complex-valued processes, for $1 \leq n \leq N$,
\begin{equation}
  u_n^M (t)
  =
  \frac{1}{\sqrt{2\pi}} \int_0^t \sum_{m=-M}^M
    \alpha_{m,n} \rme^{- \rmi (m + \sigma_n) s} \rmd s
  \label{defu}
\end{equation}
for $t \in \RR$. By construction, $u_n^M$ is analytic for any finite $n,M$, and
  $ \rme^{\rmi \sigma_n t} \, \rmd u_n^M / \rmd t $
is a family ($1 \leq n \leq N$) of independent $2\pi$-periodic complex processes. 
In the limit $M \to \infty$, the processes $u_n^M$ lose their smoothness 
(as, typically, their Fourier coefficients decay slowly), 
but Proposition~\ref{Lipv} shows that they almost surely (a.s.) admit a H\"older
continuous limit $u_n$.

Specifically, we characterize the smoothness of a function $y \in C(\RR, \CC)$ 
by its modulus of continuity \cite{Kah85,Kal01}, 
\begin{equation}
  \omega_y : ]0, \infty[ \to [0, \infty] : 
  h \mapsto \omega_y(h) = \sup_{\abs{t-t'} \leq h} \abs{y(t') - y(t)} \, .
  \label{defom}
\end{equation}

Our first objective is a gaussian convergence theorem, in the limit $N \to \infty$, for
the complex-valued process $U_N = N^{-1/2} \sum_n u_n$.  Let
\begin{equation}
  U_N^M (t)
  =
  \frac{1}{\sqrt{N}} \sum_{n=1}^N u_n^M(t)
  \label{defUNM}
\end{equation}
for $t \in \RR$. Note that if the $\sigma_n$'s do not vanish and the $\alpha_{m,n}$ are
i.i.d., processes $u_n$ are not i.i.d., but they remain independent with closely related
moments.

For $g \in C^1([0, 2\pi], \CC)$, let
%  $\hat g_m = (2\pi)^{-1/2} \int_0^{2\pi} g(t) \rme^{- \rmi m t} \rmd t$
% and
\begin{equation}
  \hat g_{m,n}
  =
  (2\pi)^{-1/2} \int_0^{2\pi}
                    g(t) \rme^{- \rmi (m + \sigma_n) t} \rmd t
  \, .
\end{equation}

We also introduce the $N$ complex-valued processes, for $1 \leq n \leq N$,
\begin{equation}
  y_n^M (t)
  =
  \frac{1}{\sqrt{2\pi}} \int_0^t \sum_{m=-M}^M
    \alpha_{m,n} \rme^{- \rmi m s} \rmd s
  \label{defv}
\end{equation}
for $t \in \RR$. In case the $\alpha_{m,n}$ are i.i.d., the processes $y_n^M$ are
i.i.d.\ for given $M$.

\begin{Prop}
  \label{contmod}
  Let $\sigma \in \RR$.
  If $y \in C(\RR, \CC)$ has a modulus of continuity $\omega_y$,
  and $u \in C(\RR, \CC)$ is defined by
  $u(t) = \int_0^t \rme^{- \rmi \sigma s} \rmd y(s)$ for $t \in \RR$,
  then its modulus of continuity satisfies
  $\omega_u(h) \leq (1 + \abs{\sigma} h) \omega_y(h)$
%  and
%  $\omega_y(h) \leq (1 + \abs{\sigma_n}h) \omega_u(h)$
  for $h \geq 0$.
\end{Prop}

\noindent \textit{Proof}
First note that, for any $t, t' \in \RR$,
\begin{eqnarray}
%  & &
  \rme^{\rmi \sigma t'} (u(t') - u(t))
  % \nonumber \\
   & = &
  \int_t^{t'} \rme^{- \rmi \sigma (s-t')} \rmd y(s)
  \nonumber \\
  & = &
  y(t') - y(t) + \int_t^{t'} (\rme^{- \rmi \sigma (s-t')} - 1) \rmd y(s)
  \nonumber \\
  & = &
  y(t') - y(t)
  + \left[ (\rme^{- \rmi \sigma (s-t')} - 1) (y(s) - y(t)) \right]_t^{t'}
  \cr
  &&  + \int_t^{t'} (y(s) - y(t)) \rmi \sigma \rme^{- \rmi \sigma (s-t')} \rmd s
\label{contmod1}
\end{eqnarray}
by integration by parts. The middle term in the right hand side of (\ref{contmod1}) vanishes, 
and we estimate
the sum using triangle inequality for $t \leq t'$,
\begin{equation}
  \abs{ u(t') - u(t) }
  =
  \abs{ \rme^{\rmi \sigma t'} (u(t') - u(t)) }
  \leq
  \abs{y(t') - y(t)}
  + \int_t^{t'} \abs{y(s) - y(t)} \abs{\sigma} \rmd s
\label{contmod2}
\end{equation}
from which the  claim follows by definition of the moduli of continuity.
  % The second claim is proved similarly.
\qed

For $0 < \beta \leq 1$ and $p \in \NN$, we denote by $C^{p,\beta}(\RR,\CC)$ the class
of continuous complex-valued functions of a real variable, 
with $p$ continuous derivatives, 
such that their $p$-th
derivative is H\"older continuous with exponent $\beta$.

\begin{Prop}
  \label{Lipv}
  Let $u_n^M$ and $y_n^M$ be defined by (\ref{defu}) and (\ref{defv})
  under assumptions (S2).
  For any $0 < \beta < 1/2$, and for any $n$,
  the sequences $y_n^M$ and $u_n^M$ converge a.s.\ in 
  $C^{0,\beta}(\RR,\CC)$
  as $M \to \infty$.
\end{Prop}

\noindent \textit{Proof}
The $y$ statement results immediately from Theorem 3, Sec.~7.4 in \cite{Kah85}, as we
compute the sums $s_j^2 = \sum_{m = 2^j}^{2^{j+1}-1} m^{-2} \EE A_{m,n}^2 \approx
\left[m^{-1}\right]_{2^j-1/2}^{2^{j+1}-1/2} \approx 2^{-j+1/2}$, using the fact that
$\EE A_{m,n}^2 = 1$.

The $u$ statement then follows from Proposition~\ref{contmod}.
  \qed

% *** comment on finite $1/\beta$-variation

\begin{Prop}
  \label{gunM}
  Let $u_n^M$ and $y_n^M$ be defined by (\ref{defu}) and (\ref{defv})
  under assumptions (S2).
  For any $g \in C^1([0, 2\pi], \RR)$, consider the complex random variables
  $(g, u_n^M) := \int_0^{2 \pi} g(t) \rmd u_n^M(t)$.
  \\
  (i) For any $M$,
  $\EE (g, u_n^M) = 0$,
  $\EE (g, u_n^M)^2 = \sum_{m=-M}^M {\hat g}_{m,n}^2 \EE \alpha_{m,n}^2$
  and
  $\EE \abs{(g, u_n^M)}^2
  = \sum_{m=-M}^M {\hat g}_{m,n}^* {\hat g}_{m,n}$.
  Moreover,
  $\sup_{n,M} \EE \abs{(g, u_n^M)}^4 \leq (2 + C_4) \norm{g}_2^4$.
  \\
  (ii) Assume further that the $\alpha_{m,n}$'s are four-symmetric.
  Then as $M \to \infty$, the complex r.v.'s
  $(g, u_n^M)$ converge a.s.\ to a r.v.\ $(g, u_n)$ such that
  $\EE (g, u_n) = 0$,  $\EE (g, u_n)^2 = 0$,
  $\EE \abs{(g, u_n)}^2 = \int_0^{2\pi} g^2(t) \rmd t$, and
  $\sup_n \EE \abs{(g, u_n)}^4 \leq (2 + C_4) \norm{g}_2^4$.
\end{Prop}

\noindent \textit{Proof}
Calculations are straightforward as the given test function $g$ is continuous and $[0,2\pi]$ is compact~:
\begin{eqnarray}
  \EE (g, u_n^M)
  & = &
  \frac{1}{\sqrt{2\pi}} \int_0^{2 \pi} g(t) \sum_{m=-M}^M
    \EE \alpha_{m,n} \rme^{- \rmi (m + \sigma_n) t} \rmd t
  =
  0 \, ,
  \\
  \EE (g, u_n^M)^2
  & = &
  \frac{1}{2\pi} \int_0^{2 \pi} \int_0^{2 \pi}
    g(t) g(s) \sum_{m=-M}^M \EE \alpha_{m,n}^2 \rme^{- \rmi (m + \sigma_n) (t+s)}
    \rmd t \rmd s
  \nonumber \\
  & = &
  \sum_{m=-M}^M \EE \alpha_{m,n}^2 \hat g_{m,n}^2 \, ,
  \\
  \EE \abs{(g, u_n^M)}^2
  & = &
  \frac{1}{2\pi} \int_0^{2 \pi} \int_0^{2 \pi}
    g(t) g(s) \sum_{m=-M}^M \EE A_{m,n}^2 \rme^{- \rmi m (t-s)}
    \rmd t \rmd s
  \nonumber \\
  & = &
  \sum_{m=-M}^M \EE A_{m,n}^2 \abs{\hat g_{m,n}}^2
  \, .
\end{eqnarray}
Given that $\EE A_{m,n}^2 = 1$, the latter expression yields\footnote{%
As pointed out by a referee, this argument reduces to Bessel's inequality, 
when one views $u_n^M$ as a sum of $2M+1$ basis functions, in the Hilbert space (whose elements are stochastic processes $u$) with scalar product $(u,v) = \EE \int_0^{2\pi} u^*(s) v(s) \rmd s$. Our assumptions on the r.v.'s $\alpha_{m,n}$ ensure orthonormality of our basis.}
 by Parseval's identity
\begin{equation}
  \EE \abs{(g, u_n^M)}^2
  \leq
  \sum_{m=-\infty}^\infty \abs{\hat g_{m,n}}^2
  =
  \int_0^{2\pi} g^2(t) \rmd t
  =
  \norm{g}_2^2
\end{equation}
with equality in the limit $M \to \infty$.
Finally,
\begin{eqnarray}
  \EE \abs{(g, u_n^M)}^4
  & = &
  \EE \sum_{m_1, m_2, m_3, m_4}
    \alpha_{m_1,n} \alpha_{m_2,n} \alpha_{m_3,n}^* \alpha_{m_4,n}^*
    \hat g_{m_1,n} \hat g_{m_2,n} \hat g_{m_3,n}^* \hat g_{m_4,n}^*
  \nonumber \\
  & = &
  \sum_m \EE A_{m,n}^4 \abs{\hat g_{m,n}}^4
  +
  2 \sum_{m_1 \neq m_2}
      \EE A_{m_1,n}^2 \EE A_{m_2,n}^2 \ \abs{\hat g_{m_1,n}}^2 \ \abs{\hat g_{m_2,n}}^2
  \nonumber \\
  & \leq &
  (C_4 + 2) \ \norm{g}_2^4 \, ,
\end{eqnarray}
where the first equality follows from the definition of Fourier coefficients $\hat
g_{m,n}$, the second equality from the known first two moments of $\alpha$, and the
final inequality from the bound $C_4$ on $\EE A^4$.
\qed

Now we can prove our main claim,
\begin{Theo}
  \label{ThmU}
  Under assumption (S4), the process $U_N^M$ defined by (\ref{defUNM})
  converges in distribution to the brownian motion
  in $C([0,2\pi],\CC)$ as $N \to \infty$ and $M \to \infty$.
\end{Theo}

\noindent \textit{Proof}
First, consider the process $U_N = \lim_{M \to \infty} U_N^M$ in $C^{0,\beta}(\RR,\CC)$
for any $0 < \beta < 1/2$. The convergence is a.s.\ since $U_N^M$ is a finite linear
combination of the processes $u_n^M$. Given any $g \in C^1([0,2\pi],\RR)$, we show below
that the r.v.\ $Z_N := (g, U_N)$ converges in distribution to a normal r.v.\ $Z = X +
\rmi Y$ with $\EE Z = 0$, $\EE X^2 = \EE Y^2 = {\frac{1}{2}} \norm{g}_2^2$, $\EE
(XY) = 0$. As $C^1$ is dense in $L^2$, the same holds true for $g \in
L^2([0,2\pi],\RR)$, which will imply that the limit $(g,U)$ is the Wiener integral
\cite{Nualart06}.

The first two moments of $Z$ follow easily from the fact that the r.v.'s $\zeta_n =
\xi_n + \rmi \eta_n := (g, u_n)$ are independent. Proposition~\ref{gunM} states that
$\EE \xi_n = \EE \eta_n = 0$, $\EE (2 \xi_n \eta_n) = \Im \EE (g, u_n)^2 = 0$ and $\EE
(\xi_n^2 - \eta_n^2) = \Re \EE (g, u_n)^2 = 0$. Besides, $\EE (\xi_n^2 + \eta_n^2) = \EE
\abs{(g, u_n)}^2 = \norm{g}_2^2$.

The fourth moment condition implies the Lindeberg condition on the sequence $\zeta_n$
(alternatively, one may adapt the standard proof of the central limit theorem 
via the characteristic function),
so that $N^{-1/2} \sum_{n=1}^N \zeta_n$ converges in distribution to a normal complex
random variable, by the gaussian convergence theorem (e.g.\ Theorem 5.12 in
\cite{Kal01}).
\qed

\begin{Rema}
  Our statement holds for arbitrary choice of coefficients $\sigma_n$, essentially
  thanks to the fact that, for any $\sigma$,
  functions $(2\pi)^{-1/2} \rme^{\rmi (m + \sigma) t}$
  form an orthonormal basis of $L^2([0,2\pi],\CC)$.
  In the special case where the $\alpha_{m,n}$ are already normal, each $u_n$ is already
  a Wiener process.
\end{Rema}

\begin{Rema}
  In the case where all $\sigma_n = 0$, the processes
  $u_n(t) - \frac{t}{2\pi} u_n(2\pi)$ define $2\pi$-periodic functions in $C^{0,\beta}(\RR,\CC)$~;
  their restrictions to $[0, 2\pi]$ converge to the brownian bridge (see \cite{Kal01}, ch.~13)
  and the $N^{-1/2} \sum_n \alpha_{m,n}$ converge to i.i.d.\ normal r.v.'s.
\end{Rema}

B\'enisti and Escande \cite{Ben95,BE97,BE98} consider the case where $\alpha_{m,n}$
is uniformly distributed on the unit circle, and work with $N=1$. 
Our statements do not formally apply to such a case. 
But they let their wave amplitude $\cA \to \infty$, so that $s \to \infty$
and $\Delta v_\rmb \to \infty$. 
To keep finite velocity and amplitude scales, 
we reformulate their case by relabeling the waves with integer-valued index $m' = m N + n$, 
letting $\sigma_n = n/N$, 
and rescaling time as $t' = t/N$, so that $m' t' = (m + \sigma_n) t$. 
To address finite $t'$ scales (of interest to them), 
we need to extend our previous statements to arbitrarily large time~; 
the following statement is a first step in this direction.

\begin{Prop}
  Under assumption (S4), assume further that
  $\sigma_n = n/N$. Then
  the process $U_N^M$ defined by (\ref{defUNM})
  converges in distribution to the brownian motion
  in $C(\RR,\CC)$ as $N \to \infty$ and $M \to \infty$.
\end{Prop}
\noindent \textit{Proof}
It suffices to prove convergence over an arbitrarily long time interval $[0, 2 \pi K]$,
with $K \in \NN_0$. To extend the previous theorem to $K > 1$, consider first a
subsequence $N = N' K$ with $N' \to \infty$. Then let $s = K s'$ and decompose $n = n' +
k N'$ with $1 \leq n' \leq N'$ and $0 \leq k \leq K-1$. Note that
\begin{eqnarray}
  U_N^M (t)
  & = &
  N^{-1/2} \sum_{n=1}^{N'K} \frac{1}{\sqrt{2\pi}}
    \int_0^{t/K} \sum_m
      \alpha_{m,n} \rme^{- \rmi (m K + k + \frac{n'}{N'}) s'} K \rmd s'
  \\
  & = &
  \sqrt{\frac{K} {N'}} \sum_{n'=1}^{N'} \frac{1}{\sqrt{2\pi}}
    \int_0^{t/K} \sum_m \ \sum_{k=0}^{K-1}
      \alpha_{m,n} \rme^{- \rmi (m K + k + \frac{n'}{N'}) s'} K \rmd s'
\end{eqnarray}
where the latter expression is equivalent to a process $\sqrt{K} U_{N'}^{MK} (t/K)$, up
to the $K-1$ terms for which $MK < mK+k \leq MK + K-1$. These $K-1$ terms do not spoil
the limit as their contribution vanishes a.s.\ for $N' \to \infty$, while by
Theorem~\ref{ThmU} the process $\sqrt{K} U_{N'}^{MK} (t/K)$ converges in distribution to
$\sqrt{K} W(t/K)$, which is distributed as $W(t)$.

Now, if $N = N' K + k$ with $1 \leq k < K$,
 the difference $U_N^M - (1 + k/N)^{-1/2} U_{N'K}^M$
converges a.s.\ to zero as $N' \to \infty$,
 while $\lim_{N' \to \infty} (1 + k/N)^{-1/2} = 1$,
so that the sequence $U_N^M$ converges like the subsequence $U_{N'K}^M$.
\qed

\begin{Rema}
  In rough paths terms (see e.g.\ \cite{LejayAnother}, sec.~8.4, 
  and \cite{FV06} for definitions and notations), 
  Theorem~\ref{ThmU} corresponds to the natural extension or lift 
  $\bfU_N$ (with $\bfU_N^1 = U_N$) 
  converging in distribution to the geometric enhanced brownian motion 
  in $C^{0,\beta}([0, 2 \pi], G^2(\CC))$ 
  for $1/3 < \beta < 1/2$.
\end{Rema}

%===========================================================
\section{Particle motion}
 \label{secParticle}
%===========================================================

We now turn to the solution of differential equations with control $U_N^M$, viz.\ to the
motion of particles in the wave field associated with the $u_n^M$'s. Since the latter
functions are $C^1$, integration against them must be interpreted so that the limit
differentials $\rmd \Re U$, $\rmd \Im U$ are the Stratonovich ones
\cite{WongZakai65,Doss77,Sussman78}. This is satisfactory for the physicists applying
e.g.\ diffusion models, but our result goes further~: this formulation opens the way to
analysing almost every single realization of the underlying noise, which need not be
gaussian.

Specifically, the motion of a particle in the prescribed field of electrostatic waves is
described by the system
\begin{eqnarray}
  \rmd q_N^M
  & =&
  \frac{\cA}{\mathfrak m} \,
    p_N^M \rmd t
  \label{dq1}   \, ,
  \\
  \rmd p_N^M
  & = &
  N^{-1/2} \sum_{n=1}^N \sum_{m=-M}^M
    A_{m,n} \sin (q_N^M(t) - (m + \sigma_n) t + \varphi_{m,n}) \rmd t
  \label{dp1}   \, ,
  \\
  & = &
  \sin (q_N^M(t)) \rmd \Re U_N^M(t) + \cos (q_N^M(t)) \rmd \Im U_N^M(t)
  \label{dp1b}   \, ,
  \\
  q_N^M(0) & = & q_0
  \, , \quad
  p_N^M(0) = p_0
  \label{pq01}   \, ,
\end{eqnarray}
where $\cA$ is an overall amplitude scale  (incorporating $\mathfrak e$) for the waves 
and $\mathfrak m$ is the particle mass. 
We rescaled the particle momentum $p$ by this overall amplitude to
construct an appropriate limit below.

\begin{Rema}
  In the special case where $\sigma_n = 0$ for all $n$, the $N$-averaging generates
  gaussian coefficients for the Fourier wave components for each $m$. In the case where
  $\sigma_n = n/N$, the wave field has period $2 \pi N$, but its sampling over the
  shorter interval $[0, 2 \pi]$ prevents the observer in the limit $N \to \infty$
  from distinguishing it from an actual white noise.
\end{Rema}

The previous section implies that, in the limit $M \to\infty, N \to \infty$, the
equations of motion may be interpreted as
\begin{eqnarray}
  \rmd Q
  & =&
  \frac{\cA}{\mathfrak m} \,
    P \rmd t
  \label{dq2}   \, ,
  \\
  \rmd P
  & = &
  \sin (Q(t)) \circ \rmd \Re U(t) + \cos (Q(t)) \circ \rmd \Im U(t)
  \label{dp2}   \, ,
  \\
  Q(0) &=& q_0
  \, , \quad
  P(0) = p_0  \, ,
  \label{pq02}
\end{eqnarray}
where $\circ \rmd$ denotes the Stratonovich differential \cite{WongZakai65}.

\begin{Prop}
  \label{WienerOnePart}
  In the limit $M \to \infty, N \to \infty$,
  the process $(q_N^M, p_N^M)$, defined by (\ref{dq1})-(\ref{dp1})-(\ref{pq01})
  under assumption (S4) with $(q_0, p_0) \in \RR^2$,
  converges in law to $(Q,P)$, where 
  \begin{eqnarray}
    Q(t) & = & q_0 + \frac{\cA}{\mathfrak m} \bigl( p_0 t + \int_0^t B(s) \rmd s \bigr), \\
    P(t) & = & p_0 + B(t) \, ,
  \end{eqnarray}
  with $B$ the standard one-dimensional Wiener process in $C([0, 2\pi], \RR)$.
\end{Prop}

\noindent \textit{Proof}
Theorem \ref{ThmU} ensures that the limit $U$ is a standard complex Wiener process.
Then, for the system (\ref{dq2})-(\ref{dp2})-(\ref{pq02}) the mapping $C([0, 2\pi],\CC)
\to C([0,2\pi],\RR^2) : U \mapsto (P,Q)$ is continuous for the topology of uniform
convergence \cite{Doss77,Sussman78}, and the continuous mapping theorem \cite{Kal01}
transfers the convergence in distribution from the control $U$ to the particle evolution
$(P,Q)$.

The proof then follows Ref.~\cite{ElPa10}. First note that the Stratonovich solution
defined by (\ref{dq2})-(\ref{dp2})-(\ref{pq02}) with the Wiener process $( \Re U(t),
\Im U(t) )$ coincides with the It\^o solution because the vector fields $\sin(q)
\partial_p$ and $\cos(q) \partial_p$ commute. Finally, since $\cos^2 q + \sin^2 q = 1$,
the process defined by $\rmd P = \sin Q \, \rmd \Re U(t) + \cos Q \, \rmd \Im U(t)$ is
distributed as the Wiener process in $C([0, 2\pi], \RR)$.
\qed

Now we turn to the limit ${\cA}/{\mathfrak m} \to \infty$. In this limit, we know that
the velocity components of the motions of $\cN$ particles also converge jointly in
distribution to $\cN$ independent Wiener processes. The previous results then imply

\begin{Theo}
  \label{WienerNPart}
  Given $\cN$ initial data $(q_0^\ell, p_0^\ell)$ in $\RR^2$ ($1 \leq \ell \leq \cN$), such that
  $1 - \cos(q_0^\ell - q_0^{\ell'}) + c \abs{p_0^\ell - p_0^{\ell'}}^2 > 0$ pairwise for some $c>0$,
  consider the resulting solutions to (\ref{dq1})-(\ref{dp1})-(\ref{pq01}).
  Then given any $K>0$, for ${\cA}/{\mathfrak m} \to \infty $,
     $N \to \infty $, $M \to \infty$,
  the $\cN$-dimensional process $p_N^M$
  converges to an $\cN$-dimensional Wiener process,
  and convergence is in law in $C([0, 2\pi K], \RR^\cN)$
  with the topology of uniform convergence.
\end{Theo}

This follows immediately from Theorem 3.1 in \cite{ElPa10}, using the brownian limit $U$ and the
continuous mapping theorem as in the proof of Proposition \ref{WienerOnePart} just given.

%===========================================================
\section{Interpretation of the results}
 \label{secInterpretation}
%===========================================================

Our formulation of the limit theorem is rather formal, 
and our proof strategy differs from the more usual ones in the physics literature.  

This paper starts by reducing ``noisy wave fields'' to ``white'' ones in the dense spectrum limit, using a \emph{central limit} theorem in the ``\emph{wave field} space'' of functions $U_N^M$, 
as shown in sec.~\ref{secWaveConv}.
Considering functions $u$ and $U$ is a way to get a handle on the limit process 
driving the particle motion, while it is harder to define directly the limit 
in terms of the noise ``$\rmd u / \rmd t$''.
The dense spectrum limit is instrumental here to provide the many independent terms in the sum defining the wave field. 

The resulting wave field entails the brownian limit for the velocity of a single particle for $0 \leq t \leq 2\pi$ \cite{ElPa10}. This single particle statement makes no reference to any velocity distribution function~: we take a ``trajectory'' viewpoint on stochastic processes, and state a ``diffusion process'' limit rather than a ``Fick equation'' limit.

The diffusion picture for $\cN$ particles also follows 
from our previous proof \cite{ElPa10} that, 
if the wave field is a ``periodic white noise'', 
then particles released in the resulting force field 
are independent in the $\cA \to \infty$ limit. 
This independence between particles results from the fact that particle velocities are a continuous martingale (viz., given the wave field history and their own, their velocity increments have vanishing conditional expectation), 
from the fact that a martingale is completely characterized by its quadratic variation process (which eliminates the need for considering more than two particles jointly), and 
from the ergodicity of the random evolution of the relative velocity of any pair of particles. Estimates in  \cite{ElPa10} are rather technical, and one may wish to revisit them to provide explicit rates of convergence.

Our order of limits is important~: 
first we take the dense spectrum limit $s \to \infty$, 
then we let $\cA \to \infty$,
and finally we consider $\cN \geq 1$ and $K \geq 1$,
for a single realization of the wave field.
Our convergence is in distribution with respect to the wave field random data.

In contrast, usual arguments for the Fokker-Planck limit invoke a loss of memory for the particle motion, directly in terms of particle velocity. The gaussianity of the velocity distribution at a time $t$ (given a Dirac at time $0$) is then seen as resulting from a central limit theorem with a sum over (\emph{time}-)successive independent increments. The quantity of interest 
(see e.g.\ eq.~(9.32) in \cite{HaWa04}) is often 
the (wave field) ensemble-\emph{averaged} 
velocity distribution function rather than the empirical distribution driven by a single wavefield.

%===========================================================
\section{Perspectives}
 \label{secPerspectives}
%===========================================================

We proved that the motion of $\cN$ particles in the field of random waves 
approaches a velocity-diffusion process in the dense spectrum limit. 
Our probabilistic proof highlights a central limit behaviour, 
while the B\'enisti-Escande proof stresses the elimination of correlations 
by appropriate changes of variables. In comparison with the latter proof, 
as well as with other derivations of the quasilinear limit, 
we show that uniformity of phases is unnecessarily strong an assumption~: 
four-symmetry (viz.\ phase distribution invariant modulo $\pi/2$) is sufficient. 
We also show that the wave amplitudes need satisfy only rather mild assumptions. 

Our proof uses the specific dispersion relation of B\'enisti and Escande, $k_m = k$ for all waves, 
and the regular spacing of phase velocities as $\sigma_n = n/N$. 
The first assumption enables the decomposition 
$\sum_m \sin (q - \omega_m t + \varphi_m) = C(t) \sin q + S(t) \cos q$ 
with coefficients $C$ and $S$ independent of $q$,
and the second one permitted to use the large body of knowledge on random Fourier series.
Relaxing these assumptions is physically desirable and will be considered in future work. 

In contrast with most earlier works in the plasma physics community, 
our formulation focuses on full particle trajectories, 
rather than one-particle distribution functions. 
In particular, the joint convergence theorem \ref{WienerNPart} 
supports the familiar picture that the evolution of the empirical distribution 
$\cN^{-1} \sum_{\ell = 1}^\cN \delta (v - p_\ell/{\mathfrak m})$
a.s.\ approaches the solution of the diffusion equation $\partial_t f - \partial_v D \partial_v f = 0$ 
for large $\cN$~: this \emph{law of large numbers}, and fluctuations around it, 
require a further limit ($\cN \to \infty$) to be discussed 
in the light of It\^o's arguments \cite{Ito83}. 
In substance, our Theorem \ref{WienerNPart} establishes for velocities 
what Lebowitz and Spohn \cite{LebowitzSpohn83} called Assumption C 
on the motion of particles in position space
in order to derive Fick's law for self-diffusion.

Another extension, in the case $\sigma_n = n/N$, 
would be to allow $K = \kappa N$ with a fixed $\kappa$ in Theorem \ref{WienerNPart}, 
for it would validate the diffusion picture for times beyond the discretization time 
$\tau_{\mathrm{disc}} := 2 \pi/(k \Delta v_\varphi) = 2 \pi N$ 
viz.\ the time scale after which the wave Fourier spectrum shows its discreteness. 
While some physical applications of the dense spectrum limit 
may be viewed as rejecting $\tau_{\mathrm{disc}}$ to infinity,
the mathematical issue is interesting because of  evidence 
that the diffusion description applies to the single-particle evolution over long times
\cite{BE97,EE03,Els10}.

%===========================================================
\section*{Acknowledgements}
 \label{secAckn}
%===========================================================

This work benefited from many discussions with D.~Escande 
and members of \'equipe turbulence plasma, with E.~Pardoux, 
and with participants to the 107th statistical mechanics conference at Rutgers. 
Stimulating comments by D.~B\'enisti and D.~Escande,
and an explanation by A.~Lejay 
are gratefully acknowledged, 
as are the careful reading and constructive comments 
by the anonymous referees.

%===========================================================
 \renewcommand{\thesection}{\Alph{section}}
 \setcounter{section}{0}
%===========================================================
\section*{Appendix}
\label{secApp}
%===========================================================

In Ref.~\cite{ElPa10} we introduced the auxiliary process $(X_t,Y_t)$, 
describing the relative position and velocity of two particles evolving 
in the same wave field. 
This process solves
\begin{eqnarray}
  \rmd X_t & = & Y_t \rmd t
  \quad \quad \quad \quad , \quad
  X_0 = x \, ,
  \\
  \rmd Y_t & = & \sin(X_t) \rmd B_t
  \quad \, \ , \quad
  Y_0 = y \, ,
\end{eqnarray}
in the state space $E = \TT \times \RR \setminus \{(0,0), (\pi,0)\}$, where $\TT =
\RR/(2\pi\ZZ)$ and $B_t$ is the standard brownian motion in $C(\RR^+, \RR)$. 
We proved
there in Proposition 5.1 that, for any $(x,y) \in E$, this process a.s.\ does not reach
the points $\{(0,0), (\pi,0)\}$ in finite time. The proof in Ref.~\cite{ElPa10} does not
identify points modulo $2 \pi$ for their $x$ component~; one can streamline it as
follows.

\begin{Prop}
  \nonumber
  For any $(x,y) \in E$, $\inf \{ t>0 : \sin^2(X_t) + Y_t^2 = 0 \} = + \infty$ a.s.,
  and $\inf \{ \theta >0 : \limsup_{t \to \theta^-} (\sin^2(X_t) + Y_t^2) = + \infty \}
       = + \infty$ a.s.
\end{Prop}

\noindent \textit{Proof}
Let $R_t = \sin^2(X_t) + Y_t^2$ and define $Z_t = \log R_t$. Denote by $\tau$ either of
these stopping times, corresponding respectively to $Z_t \to - \infty$ and $Z_t \to +
\infty$. Then It\^o calculus on $[0,\tau[$ yields
\begin{eqnarray}
  \rmd \sin^2 X_t
  & = &
  (2 \sin X_t \cos X_t) \, Y_t \rmd t
  \, , \\
  \rmd Y_t^2
  & = &
  2 Y_t \sin(X_t) \rmd B_t + \sin^2 (X_t) \rmd t
  \, , \\
  \rmd Z_t
  & = &
  \frac{2 Y_t \sin(X_t) \cos(X_t) + \sin^2(X_t)}{R_t} \rmd t
  - 2 \frac{Y_t^2 \sin^2(X_t)}{R_t^2} \rmd t
  \nonumber \\ &&
  + 2 \frac{Y_t \sin(X_t)}{R_t} \rmd B_t
  \, .
\end{eqnarray}
Noting that $2 \abs{ab} \leq a^2 + b^2$ and that $\abs{\cos x} \leq 1$ yields the
estimates $Y_t^2 \sin^2(X_t) \leq R_t^2 / 4$ and 
$2 Y_t \sin(X_t) \cos(X_t) + \sin^2(X_t) \geq - R_t$,
so that on the time interval $[0, \tau[$
\begin{equation}
  Z_t \geq Z_0 - \frac{3t}{2} + \int_0^t \varphi_s \rmd B_s
\end{equation}
where $\abs{\varphi_s} \leq 1$. This ensures that $Z_t$ is bounded from below on any
finite time interval since $B_t$ is bounded. Hence $\inf\{ t>0 : R_t = 0 \} = + \infty$
a.s.

Similar upper estimates imply
\begin{equation}
  Z_t \leq Z_0 + 2 t + \int_0^t \varphi_s \rmd B_s
\end{equation}
ensuring that $Z_t$ is bounded from above on any finite time interval. Hence $\inf \{
\theta>0 : \limsup_{t \to \theta^-} R_t = + \infty \} = + \infty$ a.s.
\qed

The second claim of the present statement does not supersede Lemma 5.4 of
Ref.~\cite{ElPa10}, which proves that $Y_t$ does a.s.\ not diverge as $t \to \infty$.
The present statement only proves that $(X_t,Y_t)$ remains in $E$ for all $t > 0$ a.s.

%===========================================================

                %%%%%%%%   BIBLIOGRAPHY  %%%%%%%%%%%

%
\end{document}